\documentclass[journal,article,submit,moreauthors]{Definitions/mdpi} 
\usepackage{color}

\renewcommand{\linenumbers}{} 

\firstpage{1} 
\pubvolume{1}
\issuenum{1}
\articlenumber{0}
\pubyear{2024}
\copyrightyear{2024}
\datereceived{ } 
\daterevised{ } 
\dateaccepted{ } 
\datepublished{ } 
\hreflink{https://doi.org/} 

\Title{Elastic polar oscillations and nuclear parameters}

\TitleCitation{Elastic polar oscillations and nuclear parameters}

\Author{Hajime Sotani $^{1,2,3,4}$\orcidA{}}

\AuthorNames{Hajime Sotani}

\AuthorCitation{Sotani, H.}

\address{%
$^{1}$ \quad Department of Mathematics and Physics, Kochi University, Kochi, 780-8520, Japan\\
$^{2}$ \quad RIKEN Center for Interdisciplinary Theoretical and Mathematical Sciences (iTHEMS), RIKEN, Wako 351-0198, Japan\\
$^{3}$ \quad Kochi University of Technology, Kochi, 780-8515, Japan\\
$^{4}$ \quad Theoretical Astrophysics, IAAT, University of T\"{u}bingen, 72076 T\"{u}bingen, Germany}

\corres{Correspondence: sotani@yukawa.kyoto-u.ac.jp}

\abstract{Owing to the elasticity of a neutron star, the shear and interface modes are excited, in addition to the fundamental, pressure, and gravity modes. While direct observations of the oscillation spectrum remain challenging, potential resonance signatures from these modes could be manifested in the precursors of short gamma-ray bursts from binary neutron star mergers. If the oscillation spectrum is observed, this is a good opportunity to extract the physical properties through the technique of asteroseismology. In particular, we focus on the crustal elasticity and examine the $s$- and $i$-mode frequencies. We find that, for a given EOS specified by the nuclear saturation parameters, the radius-scaled $s$-mode frequencies and the mass-scaled $i$-mode frequencies can be expressed as a function of the stellar compactness, independently of the stiffness of the core EOS, at least when the EOS in a higher-density region can be characterized only by the specific constant sound velocity. Then, we could derive the fitting formula for the $s_i$-mode for $i=1-4$ and $i_1$-mode frequencies as a function of the stellar compactness and the combination of the incompressibility of symmetric nuclear matter, $K_0$, and the slope parameter for the nuclear symmetry energy, $L$. If the $s$- and/or $i$-mode oscillation frequencies were observed, one may get the relation between these properties by identifying the observed frequencies with specific modes, using our fitting formula, which offers useful insights into the nuclear saturation parameters and stellar properties. By combining the other observational data, it may be possible to narrow the parameter space of EOS parameters and stellar properties.
We emphasize that this article summarizes the results presented in our previous studies and discusses their implications for neutron star seismology.}


\begin{document}



\section{Introduction}
\label{sec:introduction}

Neutron stars are one of the most suitable environments to probe physics under extreme states, where the density inside the star significantly exceeds the nuclear saturation density, and the gravitational and magnetic fields become much stronger than those observed in the Solar System~\cite{ST83,LP04}. Actually, due to a poor understanding of the properties of high-density matter, the equation of state (EOS) for neutron stars is still uncertain, and as a result, an accurate overall picture of their structure has yet to be established. Thus, the astronomical observations are crucial for revealing the EOS for neutron star matter. In fact, the discoveries of $\sim 2M_\odot$ neutron stars~\cite{D10,F21,Saffer25,Romani26} have ruled out the soft EOS, with which the expected maximum mass does not reach the observed maximum mass. Also, the observation of the gravitational waves from the neutron star binary merger, GW170817~\cite{GW170817,GW170817_multi}, gives us the upper limit of the tidal deformability for a neutron star, which leads to the constraint on the $1.4M_\odot$ neutron star radius being less than 13.6 km~\cite{Annala18}. In addition, the X-ray observations of pulse profiles from the rotating neutron stars through the Neutron Star Interior Composition Explorer (NICER) Mission enable us to estimate the mass and radius for PSR J0030+0451~\cite{Riley19,Miller19}, for PSR J0740+6620~\cite{Riley21,Miller21,Dittmann24}, for PSR J0437-4715~\cite{Choudhury24}, and for PSR J0614-3329~\cite{Mauviard25}. Neutron star models constructed with the true EOS for neutron star matter have to totally satisfy these observational constraints on neutron star mass and radius. In addition to the astronomical constraints mainly on the higher-density region, the terrestrial experiments, e.g.,~\cite{Lattimer19,SNN22,SO22,SN23}, and chiral effective field theory~\cite{Drischler21} are also important for constraining the EOS for the lower-density region.

The neutron star oscillation frequencies are another important piece of information for obtaining the neutron star properties. The frequencies observed in electromagnetic and/or gravitational waves bring us information about the neutron star. In practice, the neutron stars have many kinds of eigenfrequencies, depending on the input physics~\cite{KS99,Nils21}. Since such frequencies strongly depend on the stellar interior properties, once one can find a correlation between a specific frequency and neutron star properties, one can estimate the corresponding properties from the observed frequencies by solving an inverse problem. This technique is known as (gravitational wave) asteroseismology~\cite {AK96,AK98,KAA01}, which is a similar method to seismology on the Earth or helioseismology on the Sun. For example, by identifying the frequencies of the quasi-periodic oscillations observed from magnetar flares, the crust EOS, mass, and radius of the neutron star have been constrained, e.g.,~\cite{SW09,GNHL2011,SNIO2012,SIO2019,SKS23,Sotani24a}. Although gravitational waves from isolated neutron stars have not yet been directly detected, if observed, they may reveal the mass, radius, EOS, and/or rotational properties, e.g.,~\cite{AK96,AK98,SYMT2011,GK11,PA2012,DGKK2013,Sotani21}. Furthermore, the technique of gravitational wave asteroseismology has also been adopted even in the protoneutron stars provided via core-collapse supernova, to understand the explosion mechanism and to extract the physical properties from the gravitational waves, e.g.,~\cite{Fuller15,MRBV2018,TCPF18,STT2021,Jakobus23,SMT25,Sotani_CQG26}. In a similar way, through the radial oscillation analysis, it has been also reported that the dynamical stability of a hybrid star strongly depends on the timescale of matter changing at the hadron-quark 1st-order phase transition, which suggests the possibility of the presence of a new class of ``slow stable hybrid stars"~\cite{Pereira18,Goncalves22}.

For considering a realistic neutron star model, one has to take into account the crust elasticity~\cite{CH08}, which additionally excites the elastic oscillations, such as torsional oscillations (axial type modes) and shear ($s$-) and interface ($i$-) oscillations (polar type modes)\footnote{According to the recent study, it has been reported that the canonical $i$-mode associated with the crust-core boundary vanishes in stratified neutron stars and is replaced by compositional gravity ($g$-) modes with mixed gravity-interfacial character, driven primarily by strong buoyancy in the outer core~\cite{Gao25}.}. The effect of crustal elasticity on neutron star oscillations has been discussed in a series of papers by Bastrukov et al., which laid the foundation for their ``Nuclear Elastodynamics” series, in which they pioneered the analytical formulation of the solid crust of a neutron star as an ``elastic Fermi continuum”~\cite{Bastrukov96,Bastrukov97,Bastrukov99a,Bastrukov99b}. We note that, considering a spherically symmetric stellar model as a background, the polar (even) parity perturbations are completely decoupled from axial (odd) parity ones~\cite{Nick98}. The torsional oscillations have been relatively well studied in the context of the identification of the magnetar quasi-periodic oscillations. On the other hand, the studies about the $s$- and $i$-modes are very limited (e.g.,~\cite{Finn90,YL02,Piro05,KHA15,Sotani23,Sotani24b,KS25}) because the system to solve is more complicated, even though the polar type oscillations involving the density variations are more important in the gravitational wave astronomy. The crustal torsional oscillations are only excited in the crust (elastic) region, while the $s$- and $i$-modes oscillate not inside the crust region but in the whole region of the star, even though the $s$- and $i$-modes are excited due to the crust elasticity. So, to discuss the $s$- and $i$-modes, one cannot neglect the dependence on the core EOS (or core properties).

The $s$- and $i$-modes (as well as the fundamental ($f$-) and $g$-modes) may play an important role in the evolution of a binary neutron star. The emission of gravitational waves causes the orbital frequency to increase, eventually reaching the specific mode frequency of the neutron star in the binary system before merger. In such a moment, the corresponding mode amplitude can be enhanced owing to the energy transfer from the orbital one to the oscillation one due to tidal interactions. This phenomenon is known as tidal excitation, e.g.,~\cite{Lai94,RG94,LV19,Kuan21,Kuan25}, and it is a good opportunity to detect the gravitational waves produced by neutron star oscillations if they occur efficiently. In particular, if these phenomena occur with the crust elastic oscillations, such as the $s$- and $i$-modes, the oscillation amplitude enhanced by the tidal interaction eventually breaks the neutron star crust structure, and electromagnetic waves are simultaneously radiated as the X- and/or gamma-ray burst. This is known as resonant shattering~\cite{Tsang12,Tsang13,Gittins20,Suvorov22}. If it occurs, the precursor signal resulting from the resonant shattering would be observed before the main short gamma-ray burst corresponding to the merger of the binary neutron star, which may offer information on the stellar oscillation spectra. Since the $i$-mode frequencies are typically $\sim 10-100$ Hz~\cite{Tsang12,Tsang13}, which are lower than the $s$-mode frequencies, they easily match the orbital frequencies before merger. That is, the $i$-modes may be more relevant in the observations of resonant shattering. Observationally, the precursors have been detected $1-10$ seconds before the main flare is detected~\cite{Troja10,Coppin20}, and moreover, a candidate of quasi-periodic oscillation with $\sim 22$ Hz has also been discovered in the precursor associated with GRB 211211A~\cite{Xiao24}. Anyway, once one observes the precursor signal due to the resonant shattering and identifies it with the $i$- or $s$-mode oscillations, it helps us to extract the crustal properties~\cite{Sotani23,Sotani24b,Neill21,Neill22}.

Nevertheless, the oscillation frequencies excited in a neutron star strongly depend on the neutron star properties. Namely, without understanding such dependencies, it may be difficult to extract the physical information from the (gravitational wave) frequencies, even if they are observed. One possibility to solve this model dependence is a universal relation between the frequencies and neutron star properties, independent of the model parameters, such as the EOS. In fact, several universal relations (or empirical relations) are proposed, e.g.,~\cite{AK96,AK98,Sotani21,TL2005,Chirenti15,SK21,RS23a,Guedes24,Sotani26}\footnote{It has been shown that, when considering the extended branch of hybrid stars in the slow transformation regime, i.e., slow stable hybrid stars, some of the proposed universal relations do not hold~\cite{RS22,RS23}.}. So, in this article, to derive the fitting formula for estimating the $s$- and $i$-mode frequencies, we systematically examine the dependence of their frequencies on the neutron star properties and EOS.

This article summarizes the results presented in Refs.~\cite{Sotani23,Sotani24b,KS25} and discusses their implications for neutron star seismology. This article is organized as follows. In Sec.~\ref{sec:NS}, we describe the neutron star models focusing on the crustal region and also the shear modulus characterizing the crust's elasticity. Then, in Sec.~\ref {sec:Crust_Osci}, we discuss the behavior of the $s$- and $i$-mode frequencies and their dependence on the EOS parameters, and derive the fitting formula for expressing them. In this section, we also discuss the accuracy of the Cowling approximation, which is the approximation that the metric perturbations are neglected, in the $s$- and $i$-mode frequencies. Finally, we conclude this study in Sec.~\ref{sec:conclusion}. Unless otherwise mentioned, we adopt geometric units with $c=G=1$, where $c$ and $G$ denote the speed of light and the gravitational constant, and use the metric signature $(-,+,+,+)$.

\section{Neutron star equilibrium and shear modulus}
\label{sec:NS}

As a background model for the elastic oscillations, we consider a static, strain-free, and spherically symmetric stellar model, which is described with the metric given by
\begin{equation}
  ds^2 = -e^{2\Phi}dt^2 + e^{2\Lambda} dr^2 + r^2\left(d\theta^2 + \sin^2\theta d\phi^2\right),
\end{equation}
where $\Phi$ and $\Lambda$ are metric functions that depend only on the radial coordinate $r$, and $\Lambda$ is directly associated with the mass function $m(r)$ through $e^{-2\Lambda}=1-2m/r$. We note that the mass function is the gravitational mass enclosed within the radius $r$. The stellar equilibrium is constructed by integrating the Tolman-Oppenheimer-Volkoff (TOV) equations, adopting an appropriate EOS. To systematically examine the dependence of the elastic oscillation frequencies on the nuclear parameters, we adopt a phenomenological EOS proposed by Oyamatsu and Iida~\cite{OI03,OI07} (hereafter referred to as OI-EOS).

The bulk energy, $w$, of zero-temperature nuclear matter constructed with any EOS can be expanded in the vicinity of the saturation density, $n_0$, of the symmetric nuclear matter, whose proton number density, $n_p$, is the same as the neutron number density, $n_n$, as a function of the baryon number density, $n_{\rm b}$, and the neutron excess, $\delta$:
\begin{equation}
  w = w_0 + \frac{K_0}{2}u^2 +{\cal O}(u^3) +\left[S_0+Lu +{\cal O}(u^2)\right]\delta^2, \label{eq:w}
\end{equation}
where $u\equiv (n_{\rm b}-n_0)/(3n_0)$ and $\delta\equiv (n_n-n_p)/n_{\rm b}$. The coefficients in this expansion, $w_0$, $n_0$, $K_0$, $S_0$, and $L$, are the nuclear saturation parameters, which characterize each EOS. We note that, although we consider the expansion as in Eq.~(\ref{eq:w}) in this study, one can also expand into higher-order terms and discuss the additional nuclear saturation parameters. Additionally, since the fourth-order symmetry energy at saturation density, i.e., the term proportional to $\delta^4$, is typically much smaller than the quadratic term, the empirical parabolic approximation is known to work well around nuclear saturation density~\cite{Cai12,TT13}. Among the five expansion parameters in Eq.~(\ref{eq:w}), $w_0$, $n_0$, and $S_0$ are more constrained from the terrestrial experiments due to the saturation properties of nuclear matter, while the remaining parameters, $K_0$ and $L$, are more difficult to constrain from the experiments. This is because one has to obtain the data in a relatively wider density region around the saturation density to constrain the values of $K_0$ and $L$. The fiducial values of $K_0$ and $L$ obtained from the experiments are $K_0=240\pm 20$ MeV~\cite{Shlomo06} and $L=60\pm 20$ MeV~\cite{Vinas14,Li19}, and it is also possible to constrain them via astronomical observations, e.g.,~\cite{SIO2019,LS13,LS14,Sotani26b}.

The OI-EOS is constructed for given values of $K_0$ and $L$ so that the bulk energy becomes the form given by Eq.~(\ref{eq:w}) in the limit of $\delta\to 0$ and $n_{\rm b}\to n_0$, and the other saturation parameters are tuned to reproduce the empirical nuclear data for stable nuclei adopting a simplified version of the extended Thomas-Fermi theory~\cite{OI03,OI07}. Thus, the neutron star properties in relatively lower density regions can be well described using the OI-EOS, while the discussion in higher density regions with the OI-EOS may be insufficient. Therefore, to see the dependence on the stiffness of the EOS in a higher density region, we also consider the EOS for $\varepsilon\ge\varepsilon_t$ characterized by the sound velocity, $c_s$, assuming that $c_s=(dp/d\varepsilon)^{1/2}$ in this study, i.e.,
\begin{equation}
  p=\alpha(\varepsilon-\varepsilon_t)+p_t, \label{eq:eos_high}
\end{equation}
where $\varepsilon$ and $p$ denote the energy density and pressure; $\alpha=c_s^2$; and $\varepsilon_t$ and $p_t$ are the transition density and the pressure at $\varepsilon=\varepsilon_t$, which is connected to the OI-EOS at $\varepsilon=\varepsilon_t$. 

Based on chiral effective field theory for low-energy nuclear forces, the EOS of pure neutron matter has been precisely determined with well-quantified theoretical uncertainties up to approximately twice the saturation density~\cite{Tews13}. This provides a rigorous foundation for constructing the neutron star EOS, a framework further validated by terrestrial experiments measuring collective flows in heavy-ion collisions~\cite{Fevre16}, so that the density region up to this point is widely adopted as a well-established anchor in recent studies of the neutron star EOS~\cite{Tews19},
although heavy-ion collision data are notoriously very dependent on transport models used in the analysis.
Therefore, in this study, we assume that $\varepsilon_t$ is equal to twice the saturation density.
Considering that the transition density between the crust and core is roughly one-third to one-half of the saturation density~\cite{OI07,KP26}, the crust region is well expressed by only the OI-EOS. Hereafter, unless the value of $\alpha$ is specified, the stellar structures are constructed using the original OI-EOS by integrating the TOV equations from the center to the surface.

\begin{figure}
\centering
\begin{minipage}[b]{0.49\columnwidth}
    \centering
    \includegraphics[scale=0.5]{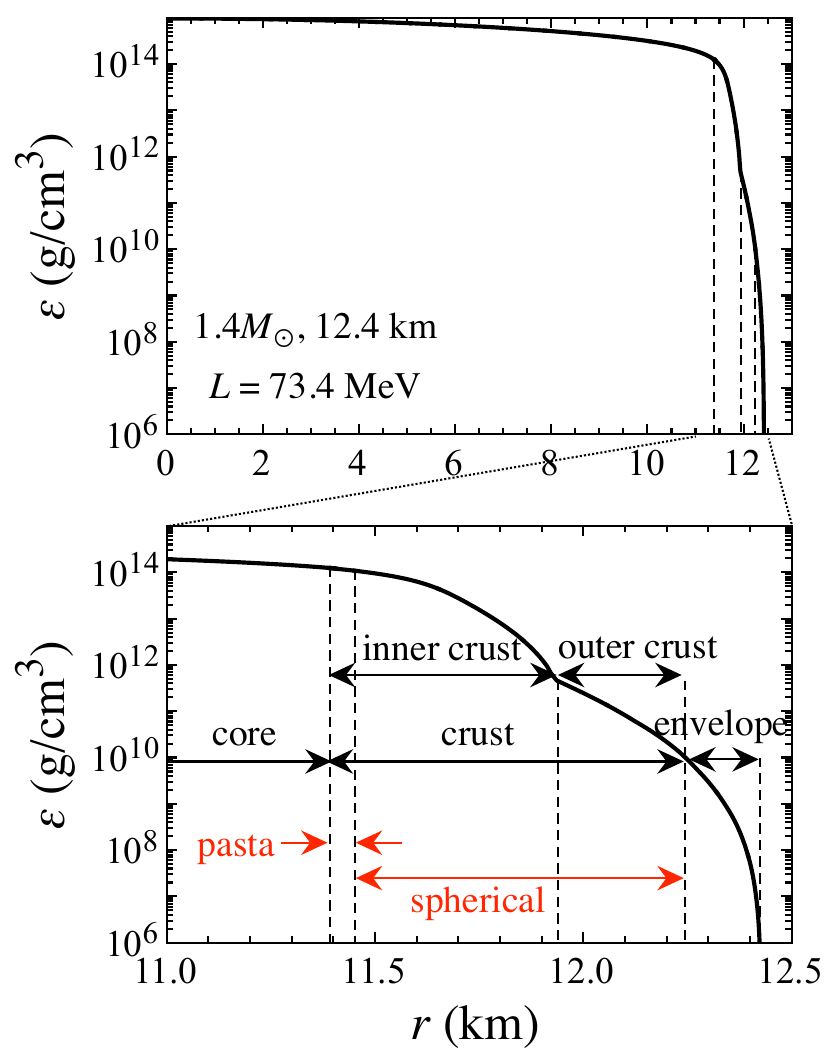}
\end{minipage}
\begin{minipage}[b]{0.49\columnwidth}
    \centering
    \includegraphics[scale=0.51]{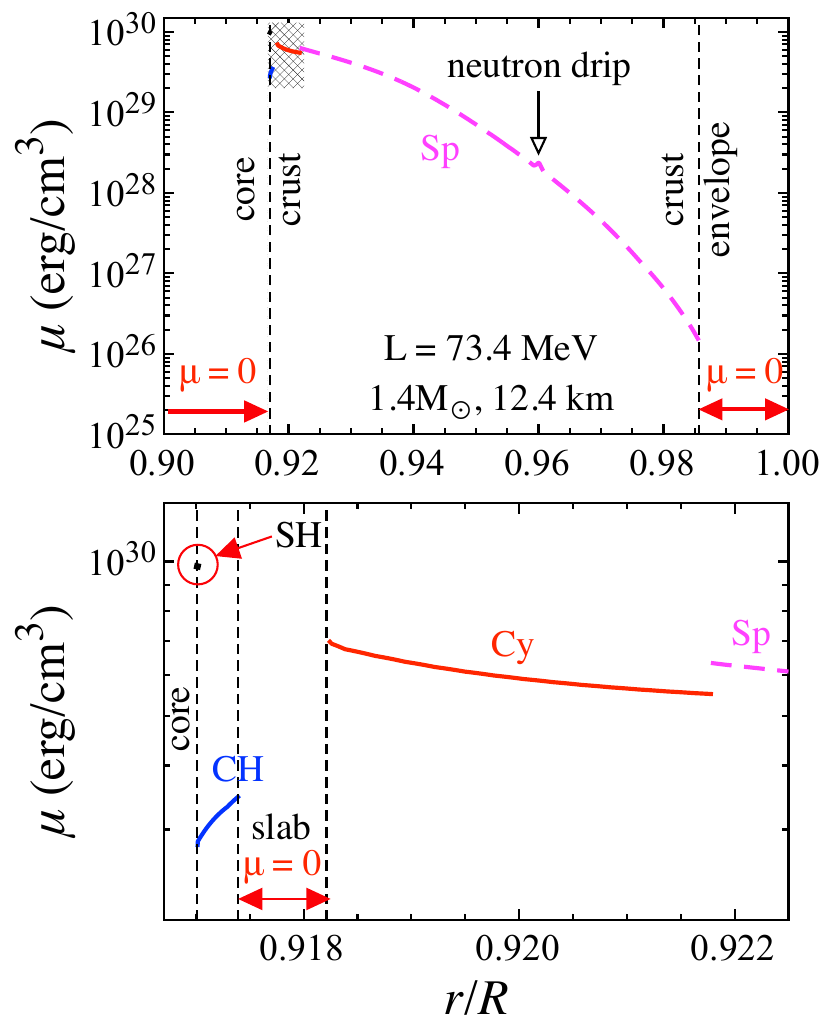}
\end{minipage}
 \caption{Left panel: Radial profile of energy density inside the neutron star with $1.4M_\odot$ and $12.4$ km, adopting the EOS with $K_0=230$ and $L=73.4$ MeV. The bottom panel is an enlarged view of the region for $r \ge 11$ km. The vertical dashed lines in the top panel denote the boundary between the core and crust, the boundary between the inner and outer crust, and the boundary between the crust and envelope, from left to right. Right panel: The radial distribution of the shear modulus, depending on the various nuclear shapes, i.e., sphere (Sp), cylindrical (Cy), cylindrical-hole (CH), and spherical-hole (SH), for the same neutron star model as on the left. The bottom panel is an enlarged view of the shaded region in the top panel.
 Taken from Ref.~\cite{Sotani23}. }
\label{fig:mu}
\end{figure}

The OI-EOS tells us the transition density from the crust to the core (and also the transition densities where the various nuclear pasta structures appear), which mainly depends on $L$, i.e., the transition density to the core region becomes larger as $L$ decreases~\cite{OI03,OI07}. The thickness of the pasta region also becomes thicker as $L$ decreases, while the pasta structure disappears when $L\gtrsim 100$ MeV~\cite{OI07}. On the other hand, the determination of the density at the crust outer surface, i.e., the transition density between the crust and envelope, is more difficult~\cite{CH08,GPE83,Hernquist84,Beznogov21}. This is because one has to take into account the physical temperature effect on the matter properties when the density becomes much lower than the saturation density. Namely, one can generally consider a zero-temperature neutron star model even though the physical temperature is around $10^8$ K, because the Fermi temperature is much higher than the physical temperature inside the neutron star. Meanwhile, in the vicinity of the stellar surface, where the density becomes much lower than the nuclear saturation density, the Fermi temperature also becomes low, and one cannot neglect the effect of the temperature. In practice, the density of the crust outer surface becomes $\sim 10^6-10^{10}$ g/cm$^3$~\cite{GPE83,Epstein88}. In this study, we simply assume that this transition density is $10^{10}$ g/cm$^3$, while the stellar surface density is $10^6$ g/cm$^3$.  How the $i$-mode frequencies shift due to the change in the transition density between the crust and the envelope is discussed in Appendix~\ref{sec:appendix_B}, using a simple model. In the left panel of Fig.~\ref{fig:mu}, as an example, we show the radial profile of the energy density for the neutron star model with $1.4M_\odot$ and 12.4 km, adopting the OI-EOS with $K_0=230$ and $L=73.4$ MeV. From this figure, one observes that the crust thickness is very tiny compared to the stellar radius, and also that most of the crust region is composed of spherical nuclei, and the pasta layer is very narrow.

The crust elasticity is characterized by the shear modulus, $\mu$. In a phase consisting of spherical nuclei arranged in a body-centered cubic (bcc) lattice, which occupies most of the crust region, the shear modulus, $\mu_{\rm sp}$, is formulated as
\begin{equation}
  \mu_{\rm sp} = 0.1194 \frac{n_i(Ze)^2}{a}, \label{eq:mu_sp}
\end{equation}
where $n_i$, $Z$, and $a$ denote the ion number density, the ion charge number, and the Wigner-Seitz cell radius, i.e., $n_i=(4\pi a^3/3)^{-1}$~\cite{Ogata90,mu_sp}. This shear modulus has been derived under the assumption that the ion is a point particle. On the other hand, although the study of elasticity in the phases of cylindrical and slab-like nuclei may be poor, as per the study by Pethick and Potekhin~\cite{PP98}, the shear modulus in the phase of cylindrical phase is expressed as
\begin{equation}
  \mu_{\rm cy} = \frac{2}{3} E_{\rm Coul}\times 10^{2.1(w_2-0.3)}, \label{eq:mu_cy}
\end{equation}
using the Coulomb energy per volume of a Wigner-Seitz cell, $E_{\rm Coul}$, and the volume fraction of cylindrical nuclei, $w_2$, while the phase of slab-like nuclei behaves as a fluid against the linear response, i.e., the shear modulus in this phase is considered as
\begin{equation}
  \mu_{\rm sl} = 0, \label{eq:mu_sl}
\end{equation}
when one considers the linear analysis. Furthermore, the shear moduli inside the phases of cylindrical-hole nuclei and spherical-hole nuclei are estimated with the formulae given by Eqs.~(\ref{eq:mu_cy}) and (\ref{eq:mu_sp}), because the structure is the same as the phases of cylindrical and spherical nuclei, respectively. Then, as shown in the right panel of Fig.~\ref{fig:mu}, one can derive the shear moduli inside the crust region, depending on the various phases. That is, the crustal structure consists of two elastic regions separated by a fluid phase composed of slab-like nuclei~\cite{SIO2019}.

Before we start to discuss the oscillation frequencies, we have to emphasize the limitation of our EOS models. The nuclear saturation parameters, which are the lowest order of the expansion coefficients, $w_0$, $n_0$, $K_0$, $S_0$, and $L$, characterize a (zero-temperature) homogeneous nuclear matter around the saturation point of symmetric matter, while the neutron star crust contains highly asymmetric and inhomogeneous matter over a broad range of sub-saturation densities. Their composition, crust-core transition, and elastic properties depend not only on the nuclear saturation parameters but also on higher-order density and isospin dependences, nuclear surface properties, and other finite-size contributions. Although the variations of these additional properties are likely to be correlated with variations of the nuclear saturation parameters within a particular nuclear parametrization, such correlations are not necessarily preserved when the additional nuclear-functional parameters are independently varied, e.g.,~\cite{Klausner26}. A similar issue arises because nuclear matter parameters alone do not fix the EOS across the full density range. In practice, how nuclear-functional uncertainties can be propagated beyond a restricted parametrization is discussed, e.g.,~\cite{Montefusco26},
which develops a flexible nuclear EOS that allows fixing a broad range of nuclear matter parameters, with a consistent treatment of the crust and core. That is, a different nuclear EOS with comparable or even identical values of $K_0$ and $L$ can produce very different stellar properties, and therefore mode frequencies.
That is, the small residuals of the fitting formulae derived within the OI family considered in this article should not be interpreted as demonstrating comparably small uncertainties in the nuclear parameters inferred from future observations. Considering these points, since we consider EOS models characterized by the nuclear saturation parameters in this article, our statement is valid only when the correlations with the nuclear saturation parameters are well preserved. If not, additional analysis is needed to validate our statement.

\section{Oscillations excited by the crust elasticity}
\label{sec:Crust_Osci}

To determine the oscillation frequencies, we simply assume the Cowling approximation in this study. That is, we consider the fluid motion on the fixed metric. Even with such an approximation, one can qualitatively discuss the behavior of frequencies~\cite{McDermott88,MK97}. In this case, the perturbation equations are derived from the linearized energy-momentum conservation law, i.e., $\delta\left(\nabla_\mu T^{\mu\nu}\right)=\nabla_\mu \delta T^{\mu\nu} =0$, where the first equality holds because of the Cowling approximation. Then, by imposing the appropriate boundary conditions at the center and surface and the junction conditions at the interfaces between the elastic and anelastic regions\footnote{As mentioned in the previous section, we construct the equilibrium model by using the EOS given by Eq. (\ref{eq:eos_high}) connected to the OI-EOS at $\varepsilon=\varepsilon_t$. So, in practice, we integrate the perturbation equations with the EOS given by Eq. (\ref{eq:eos_high}) from the center to the radial position where $\varepsilon=\varepsilon_t$, and continue integrating the perturbation equations outward with the OI-EOS.}, 
the problem to solve becomes an eigenvalue problem with respect to the eigenvalue $\omega$. Once $\omega$ is determined, the eigenfrequencies, $f$, are obtained from the resultant $\omega$ via $f=\omega / (2\pi)$. The concrete perturbation equations, the imposed boundary conditions, and the imposed junction conditions are shown in Refs.~\cite{YL02,Sotani23}. We note that one can obtain all eigenfrequencies, such as the $f$-, pressure ($p$-), $s$-, and $i$-modes, excited in the current model from the same system of equations. We note that the $g$-modes cannot be excited in our model due to the assumption of $c_s^2=dp/d\varepsilon$. In this study, we particularly focus only on the $\ell=2$ modes, because they are dominant modes in gravitational radiation~\cite{KS99,Kip69b}.

We note that if some of the dripped neutrons behave as a superfluid, which does not contribute to the effective inertia participating in the elastic oscillations, the $s$- and $i$- frequencies may increase compared to those without such effects, analogous to the crustal torsional oscillations~\cite{SNIO2013}. In addition, as mentioned in the following subsection~\ref{sec:behavior}, the number of the $i$ modes excited in the neutron star without the composition gradient is equivalent to the number of interfaces where the elasticity becomes zero~\cite{Piro05}. So, if $\mu_{\rm sl}$ has a tiny value instead of zero, different from our simple approximation, the number of interfaces changes from 4 to 2. In such a case, the $i$-modes are excited only at the boundaries between the crust and envelope and between the crust and core. Furthermore, in such a case, the junction conditions adopted in the integration of the perturbation equations are also changed, so that the $s$-mode frequencies may also be shifted.

Furthermore, recent work by Gao et al.~\cite{Gao25} has shown that, in neutron star models with stratification, the standard crust-core interface modes disappear and are replaced by modes in which gravitational and interface properties are intermingled. In other words, the mode spectrum itself can change as a result of the layered structure. On the other hand, this paper assumes that $c_s^2=dp/d\varepsilon$; as a result, the $g$ mode arising from composition no longer exists. It should therefore be noted that if composition gradients that induce the $g$-mode oscillations are taken into account, the $i$-mode discussed here may disappear.

\subsection{Behavior of the $s$- and $i$-modes}
\label{sec:behavior}

\begin{figure}[tbp]
\begin{center}
\includegraphics[scale=0.5]{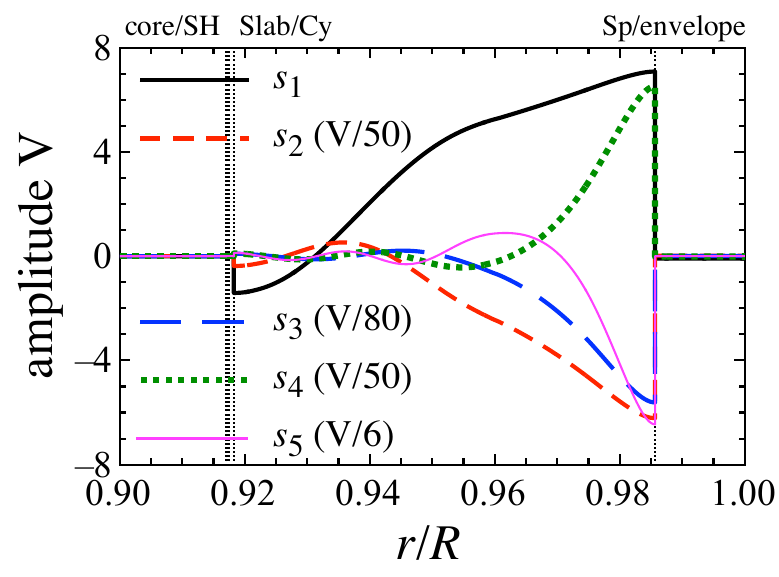} 
\end{center}
\caption{
Radial profile of the eigenfunction of the Lagrangian displacement in the angular direction, $V$, for the $s_i$-modes with $i=1-5$ for the neutron star model with $1.4M_\odot$ and 12.4 km, adopting the EOS with $K_0=230$ and $L=73.4$ MeV. The vertical lines from right to left denote the boundary between the crust and the envelope (Sp/envelope), the boundary between the phases composed of cylindrical and slab-like nuclei (Slab/Cy), the boundary between the phases of slab-like and cylindrical-hole nuclei (CH/Slab), and the boundary between the phase composed of spherical-hole nuclei and the core (core/SH).
Taken from Ref.~\cite{Sotani23}.  }
\label{fig:sv_mode}
\end{figure}

Before discussing the fitting formula, we show the behavior of the eigenfunctions of the $s$- and $i$-mode oscillations. Both modes are excited by the elasticity inside the star. The number of $s$-modes is infinite, just like $p$-modes. The lowest mode has one node in the eigenfunction, and the nodal number increases as the mode frequency becomes higher. Using the nodal number, $i$, the mode is identified as $s_i$-mode. In Fig.~\ref{fig:sv_mode}, the eigenfunction of the Lagrangian displacement in the angular direction, $V$, for the $s_i$-modes with $i=1-5$ are shown as a function of the normalized radial coordinate, $r/R$, focusing on the crust region. From this figure, one observes that the $s$-mode oscillations are almost confined inside the elastic region. At the boundary between the anelastic and elastic regions, $V$ becomes discontinuous, while the Lagrangian displacement in the radial direction, $W$, becomes continuous owing to the junction condition at the same boundary. Since the $s_i$-modes with $i$ nodes are confined inside the elastic region, one can roughly estimate the wavelength of the $s_i$-mode ($\lambda_i$) with the thickness of the elastic region, $\Delta R$:
\begin{equation}
  \lambda_i \approx 2\Delta R / i. \label{eq:lambda_i}
\end{equation}
Therefore, the frequency of the $s_i$-mode is also estimated as
\begin{equation}
  f_{s_i} \approx v_s/\lambda_i, \label{eq:f_si}
\end{equation}
where $v_s$ denotes the shear velocity, i.e., the propagation speed of elastic oscillations. Since we only focus on the frequency region less than a few kHz in this study, we can only find the $s$-mode oscillations excited in the phase of spherical and cylindrical nuclei, as shown in Fig.~\ref{fig:sv_mode}. In addition to the $s$-mode we found, the $s$-mode can also be excited inside the phase of cylindrical-hole and spherical-hole nuclei, which is another elastic region inside the star. That may be so, but the thickness of the phase of cylindrical-hole and spherical-hole nuclei, $\Delta R_{\rm CHSH}$, is much smaller than that of the phase of spherical and cylindrical nuclei, $\Delta R_{\rm SpCy}$, e.g., $\Delta R_{\rm CHSH}/\Delta R_{\rm SpCy}\lesssim 10^{-2}$. That is, as per the above rough estimation, the $s$-mode frequencies excited in $\Delta R_{\rm CHSH}$, $f_{s}^{(\rm CHSH)}$, become much higher than the $s$-mode frequencies excited in $\Delta R_{\rm SpCy}$, $f_{s}^{(\rm SpCy)}$, e.g., $f_{s}^{(\rm CHSH)}$ may be higher than $\sim 50$ kHz (likely of no interest to current gravitational wave detectors), considering that $f_{s}^{(\rm SpCy)}$ is typically higher than $\sim 500$ Hz.

\begin{figure}[tbp]
\begin{center}
\includegraphics[scale=0.5]{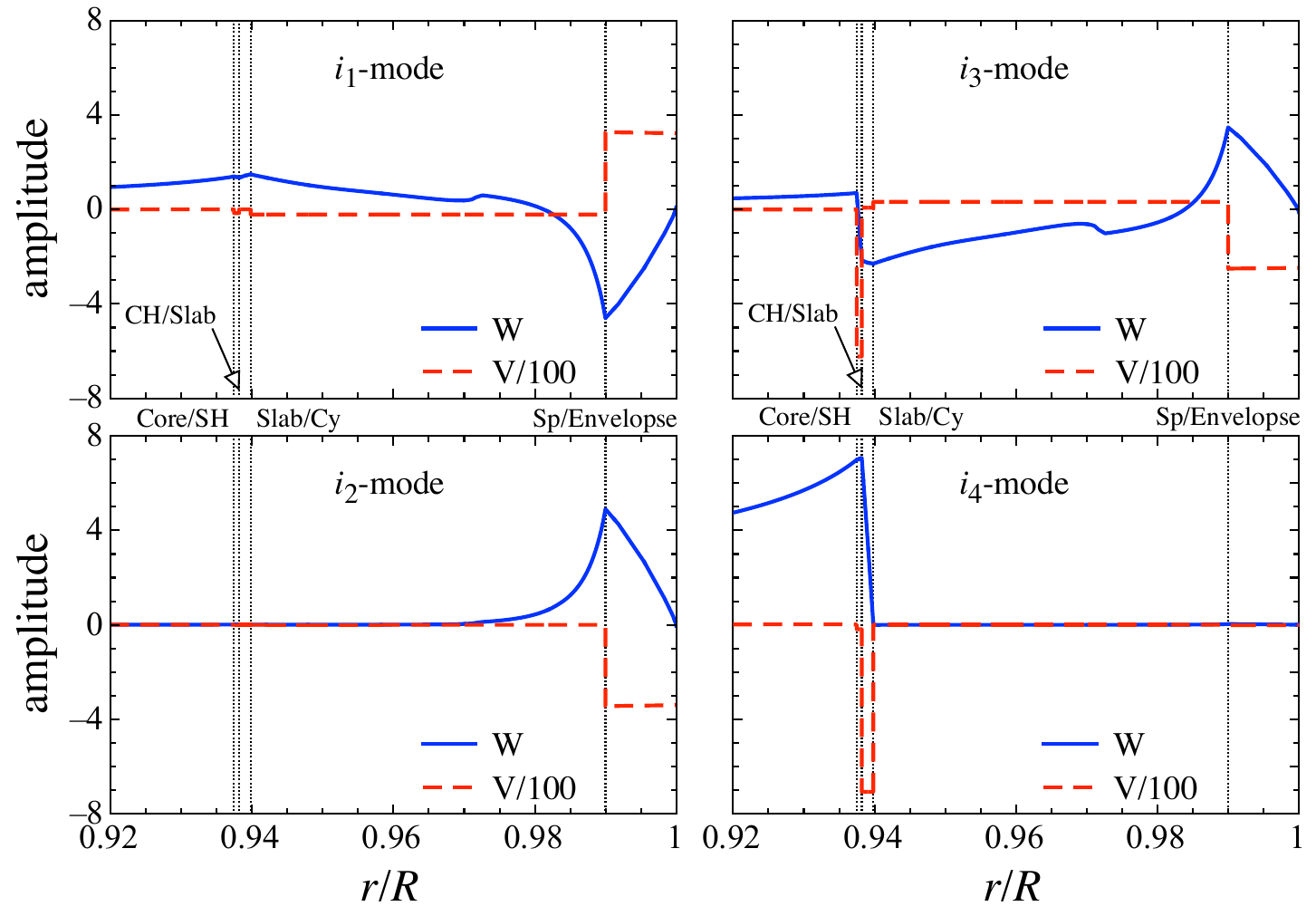} 
\end{center}
\caption{
Radial profile of the eigenfunctions of $W$ and $V$ for the $i_i$-modes with $i=1-4$ for the neutron star model with $1.44M_\odot$ and 10.2 km, adopting the EOS with $K_0=230$ and $L=42.6$ MeV. The meaning of the vertical lines is the same as in Fig.~\ref{fig:sv_mode}. Taken from Ref.~\cite{Sotani24b}. }
\label{fig:i_mode}
\end{figure}

Unlike the $s$-modes, the number of the $i$-modes is finite. How many of the $i$-modes can be excited in a neutron star corresponds to the number of interfaces between the phases with zero and nonzero elasticity~\cite{Piro05}. Since the stellar model considered in this study has four interfaces, i.e., between the envelope and the phase of spherical nuclei, between the phases of cylindrical and slab-like nuclei, between the phases of slab-like and cylindrical-hole nuclei, and between the phase spherical-hole nuclei and core region, four interface modes can be excited\footnote{In the case that the elastic region is too narrow, the number of the excited $i$-modes becomes less than the number of the interface inside the star, or some of the $i$-mode frequencies become too low to numerically determine~\cite{Sotani23}}. In addition, unlike the $s$-modes, 
since there is no established convention for numbering the subscript of the $i_i$-modes, we will number them here in order of decreasing frequency, starting with $i=1, 2, \cdots$. As an example, Fig.~\ref{fig:i_mode} shows the radial profile of the amplitude of $W$ and $V$ for the $i_i$-mode oscillations with $i=1-4$, where the vertical lines denote the interfaces between the elastic and anelastic regions. From the figure, one can observe that, just like in the $s$-mode oscillations,  due to the junction condition at the interface, $W$ is continuous while $V$ is discontinuous at the interfaces. The amplitude of $W$ increases significantly at the interface, while $V$ is almost behaves like a step function that is discontinuous at the interface. In addition, unlike the $s$-modes, the amplitude of the $i$-mode in the anelastic region becomes comparable to that in the elastic region.

\subsection{Fitting formula}
\label{sec:fit}

The torsional oscillations, which are the axial-type oscillations excited by the presence of elasticity, do not involve the stellar deformation, and they are confined inside the elastic region. So, one can discuss the frequencies of the torsional oscillations independently of the core region once one can construct the equilibrium model of the neutron star crust. On the other hand, the $s$- and $i$-modes are also the oscillation modes excited by the elasticity, but they are the polar type oscillations, which involve the radial motion. Thus, for discussing the $s$- and $i$-mode oscillations, one has to take into account not only the crust properties but also the core properties. In fact, even if one adopts the EOS with the same saturation parameters, i.e., the lower density region is described in the same way, the frequencies of the polar oscillations are shifted, depending on the core EOS (or core properties). To see such a dependence of the frequencies on the core properties, as mentioned in \S~\ref{sec:NS}, we examine the frequencies by changing the stiffness of the core EOS, characterizing its sound velocity, as in Eq.~(\ref{eq:eos_high}). 
As an example, we consider the stellar models constructed with the OI-EOS with $L=42.6$ MeV and $73.4$ MeV, fixing $K_0=230$ MeV, and those with the same OI-EOS connected to the EOS given by Eq.~(\ref{eq:eos_high}) with $\alpha=1/3$, 0.6, and 1. The mass and radius of the resultant neutron stars are shown in Appendix~\ref{sec:appendix_A}.

\begin{figure}
\centering
\begin{minipage}[b]{0.49\columnwidth}
    \centering
    \includegraphics[scale=0.5]{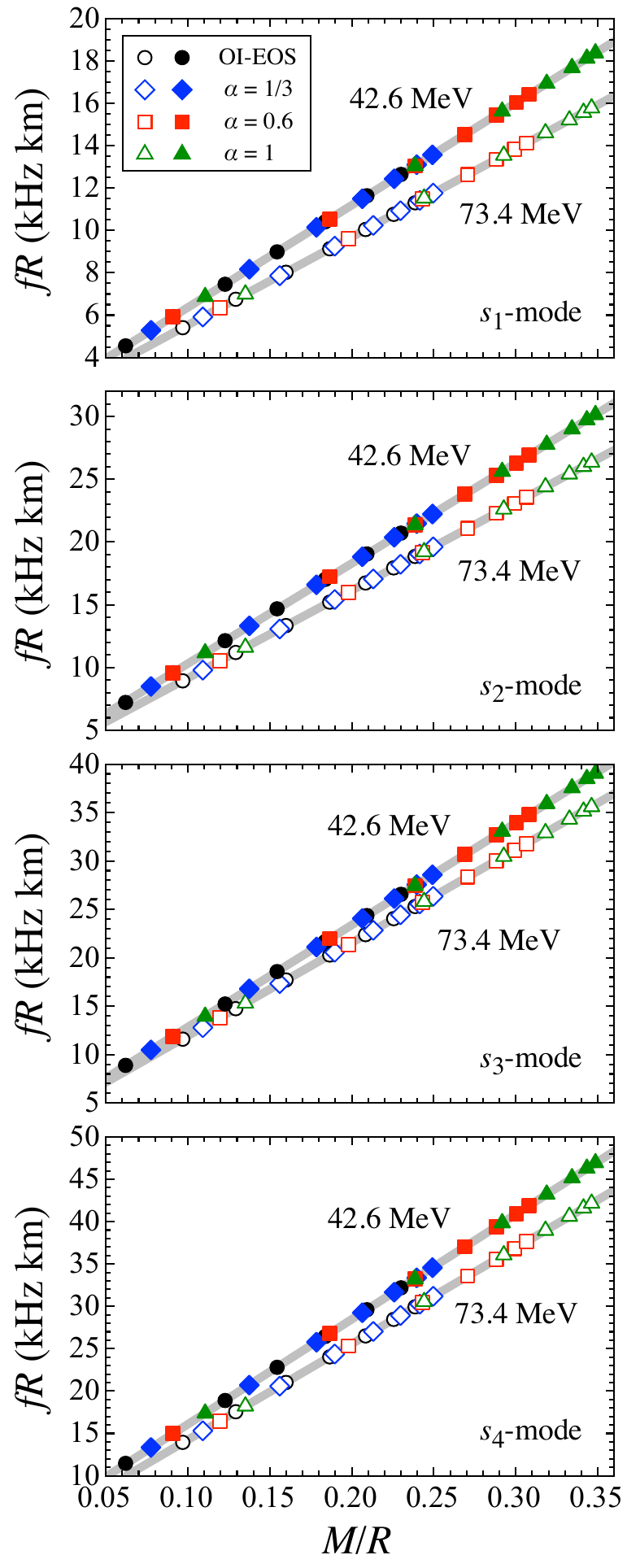}
\end{minipage}
\begin{minipage}[b]{0.49\columnwidth}
    \centering
    \includegraphics[scale=0.5]{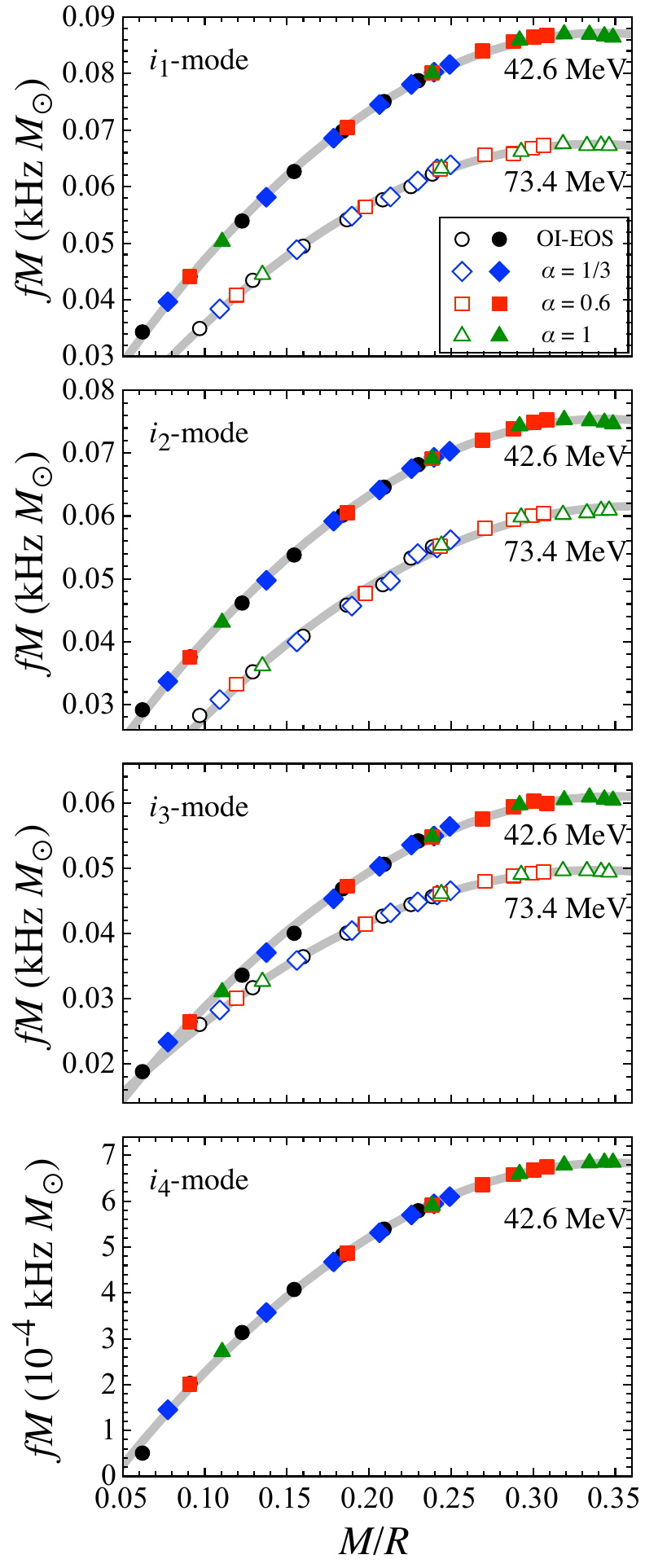}
\end{minipage}
\caption{For neutron star models constructed using the EOS with $L=42.6$ and 73.4 MeV, fixing $K_0=230$ MeV, and with different core EOS stiffness, $\alpha= c_s^2$, the radius-scaled $s_i$-mode frequencies in the left panels and the mass-scaled $i_i$-mode frequencies in the right panels are shown as a function of the stellar compactness, $M/R$. The thick solid lines denote the fitting formulae given by Eq.~(\ref{eq:sR}) in the left panel and Eq.~(\ref{eq:iM}) in the right panel.
 Taken from Ref.~\cite{Sotani23}. }
\label{fig:ismode}
\end{figure}

Then, as shown in Fig.~\ref{fig:ismode}, we find that the radius-scaled $s_i$-mode frequencies, $f_{s_i}R$, and the mass-scaled $i_i$-mode frequencies, $f_{i_i}M$, are well expressed as a function of the stellar compactness, independently of the stiffness of the core EOS, at least when the core EOS can be characterized only by the specific constant sound velocity. From this figure, we can derive the fitting formulae for $f_{s_i}R$ and $f_{i_i}M$ as a function of $\chi\equiv M_{1.4}/R_{12}$ with the normalized mass, $M_{1.4}\equiv M/(1.4M_\odot)$, and radius, $R_{12}\equiv R/ (12\ {\rm km})$, such as
\begin{eqnarray}
  f_{s_i}R\ ({\rm kHz\ km}) &=& a_{0i} + a_{1i}\chi, \label{eq:sR} \\
  f_{i_i}M\ ({\rm kHz}\ M_\odot) &=& b_{0i} + b_{1i} \chi + b_{2i} \chi^2, \label{eq:iM}
\end{eqnarray}
where $a_0$, $a_1$, $b_0$, $b_1$, and $b_2$ are adjusting coefficients, depending on the saturation parameters. This result suggests that one may be able to evaluate the nuclear saturation parameters from observations of $s$- and/or $i$-mode frequencies, together with an additional observational constraint on the neutron star mass and radius, even without knowledge of the core EOS.

Furthermore, we attempt to express the adjusting coefficients in Eqs.~(\ref{eq:sR}) and (\ref{eq:iM}) as a function of the saturation parameters, which partially succeeds~\cite{Sotani24b}. To derive such a fitting formula for the $s$- and $i$-mode frequencies, we examine the dependence of the adjusting coefficients on the saturation parameters, adopting the OI-EOS with 6 different sets of $K_0$ and $L$, i.e., $(K_0,L)=(180, 31.0)$, $(180, 52.2)$, $(230,23.7)$, $(230,42.6)$, $(230,73.4)$, and $(360,40.9)$ in the unit of MeV. Then, after much trial and error, we find that the adjusting coefficients in Eq.~(\ref{eq:sR}) can be well expressed as a function of the specific combination of the saturation parameters, $K_0$ and $L$, expressed as $y_{0i}$ and $y_{1i}$, which depend on the nodal number $i$ for the $s_i$-modes, as
\begin{gather}
  a_{0i} = a_{0i}^{(0)} + a_{0i}^{(1)} \left(\frac{y_{0i}}{100\ {\rm MeV}}\right) 
     + a_{0i}^{(2)} \left(\frac{y_{0i}}{100\ {\rm MeV}}\right)^2, \label{eq:a0i} \\
  a_{1i} = a_{1i}^{(0)} + a_{1i}^{(1)} \left(\frac{y_{1i}}{100\ {\rm MeV}}\right)
     + a_{1i}^{(2)} \left(\frac{y_{1i}}{100\ {\rm MeV}}\right)^2, \label{eq:a1i}
\end{gather}
where $y_{0i}$ and $y_{1i}$ are listed in Table~\ref{tab:yzi}, and the coefficients in Eqs.~(\ref{eq:a0i}) and (\ref{eq:a1i}), i.e., $a_{0i}^{(j)}$ and $a_{1i}^{(j)}$ for $j=0,1,2$, are listed in  Table~\ref{tab:a_01i^j}. In a similar way, the adjusting coefficients in Eq.~(\ref{eq:iM}) are also fitted as a function of the specific combination of $K_0$ and $L$ expressed as 
\begin{gather}
  b_{ji} = b_{ji}^{(0)} + b_{ji}^{(1)} \left(\frac{z_{ji}}{100\ {\rm MeV}}\right) 
     + b_{ji}^{(2)} \left(\frac{z_{ji}}{100\ {\rm MeV}}\right)^2, \label{eq:bji} \\
  b_{j4} = b_{j4}^{(0)} + b_{j4}^{(1)} \left(\frac{z_{j4}}{100\ {\rm MeV}}\right)^{-1}
     + b_{j4}^{(2)} \left(\frac{z_{j4}}{100\ {\rm MeV}}\right)^{-2}, \label{eq:bj4} 
\end{gather}
for $j=0,1,2$ and $i=1,2,3$, where $z_{ji}$ denotes $L$ or the combination of $K_0$ and $L$ depending on $j$ and $i$, which are listed in Table~\ref{tab:yzi}, even though the fitting for the $i_3$- and $i_4$-mode frequencies does not work well, as seen below, and the coefficinets in Eqs.~(\ref{eq:bji}) and (\ref{eq:bj4}), i.e., $b_{ji}^{(k)}$ for $j=0,1,2$ and $k=0,1,2$, are listed in Table~\ref{tab:b_ji^k}.

\begin{table}[tbp]
\centering
\caption{Specific combinations of $K_0$ and $L$ ($y_{0i}$ and $y_{1i}$) in the fitting of $a_{0i}$ and $a_{1i}$ expressed by Eqs. (\ref{eq:a0i}) and (\ref{eq:a1i}), and the specific combinations of $K_0$ and $L$ ($z_{ji}$ for $i=1,2,3,4$ and $j=0,1,2$) in the fitting of $b_{0i}$, $b_{1i}$, and $b_{2i}$, given by Eqs. (\ref{eq:bji}) and (\ref{eq:bj4}). }
\begin{tabular}{c|cc|ccc}
\hline\hline
       $i$           & $y_{0i}$ & $y_{1i}$ & $z_{0i}$ & $z_{1i}$ & $z_{2i}$    \\
\hline
   $1$  &  $(K_0^4L^5)^{1/9}$  &  $(K_0^5L^6)^{1/11}$ &  $(K_0^2L^3)^{1/5}$  &  $L$ & $L$  \\
   $2$  &  $(K_0L^6)^{1/7}$   & $(K_0L)^{1/2}$ &  $(K_0^3L)^{1/4}$   & $(K_0^4L)^{1/5}$  & $(K_0^9L)^{1/10}$  \\
   $3$  &  $(K_0L^3)^{1/4}$   & $(K_0L)^{1/2}$ &  $L$   & $L$  & $L$  \\
   $4$  &  $(K_0^4L^5)^{1/9}$   & $(K_0L)^{1/2}$ &  $(K_0^3L)^{1/4}$   & $(K_0^2 L)^{1/3}$  & $(K_0^2 L)^{1/3}$  \\
\hline\hline
\end{tabular}
\label{tab:yzi}
\end{table}

\begin{table}[tbp]
\centering
\caption{Coefficients $a_{0i}^{(j)}$ and $a_{1i}^{(j)}$ for $j=0,1,2$ in the fitting of $a_{0i}$ and $a_{1i}$ expressed by Eqs. (\ref{eq:a0i}) and (\ref{eq:a1i}). }
\begin{tabular}{ccccccc}
\hline\hline
         $i$         & $a_{0i}^{(0)}$ & $a_{0i}^{(1)}$  & $a_{0i}^{(2)}$ & $a_{1i}^{(0)}$ & $a_{1i}^{(1)}$ & $a_{1i}^{(2)}$    \\
\hline
   $1$  & $1.8624$ & $-0.2316$ & $-0.1364$ & $9.4614$ & $0.7380$ & $-2.1114$   \\
   $2$  & $1.5724$ & $2.5259$ & $-2.1216$ & $17.8859$ & $-3.0316$ & $-1.0974$   \\
   $3$  & $3.8459$ & $-3.8269$ & $2.3209$ & $18.4672$ & $3.3390$ & $-3.6406$   \\
   $4$  & $3.5630$ & $2.0388$ & $-2.1046$ & $28.7034$ & $-8.3357$ & $1.0013$   \\
   \hline\hline
\end{tabular}
\label{tab:a_01i^j}
\end{table}

\begin{table*}[tbp]
\centering
\caption{Coefficients $b_{ji}^{(k)}$ for $j=0,1,2$ and $k=0,1,2$ in the fitting of $b_{0i}$, $b_{1i}$, and $b_{2i}$ with Eqs.~(\ref{eq:bji}) and (\ref{eq:bj4}). }
\resizebox{\textwidth}{!}{
\begin{tabular}{cccccccccc}
\hline\hline
         $i$         & $b_{0i}^{(0)}$ & $b_{0i}^{(1)}$  & $b_{0i}^{(2)}$ & 
                          $b_{1i}^{(0)}$ & $b_{1i}^{(1)}$ & $b_{1i}^{(2)}$  &
                          $b_{2i}^{(0)}$ & $b_{2i}^{(1)}$ & $b_{2i}^{(2)}$  \\
\hline
   $1$  & $0.0038337$ & $0.0073437$ & $-0.0068798$ & $0.083392$ & $-0.042113$ & $0.024719$ 
           & $-0.020817$ & $0.010515$ & $-0.0069116$   \\
   $2$  & $0.016178$ & $-0.019248$ & $0.0048505$ & $0.048374$ & $0.010393$ & $-0.0017163$ 
           & $-0.0088285$ & $-0.0040927$ & $0.00056104$   \\
   $3$  & $0.02413$ & $-0.053581$ & $0.033518$ & $0.1714$ & $-0.2740$ & $0.1446$ 
           & $-0.040238$ & $0.059085$ & $-0.028536$   \\
   $4$  & $0.0017045$ & $-0.0071152$ & $0.0062548$ & $0.018248$ & $-0.049949$ & $0.034971$
           & $-0.0045017$ & $0.012316$ & $-0.0086167$   \\
   \hline\hline
\end{tabular}
}
\label{tab:b_ji^k}
\end{table*}

\begin{figure}[tbp]
\begin{center}
\includegraphics[scale=0.5]{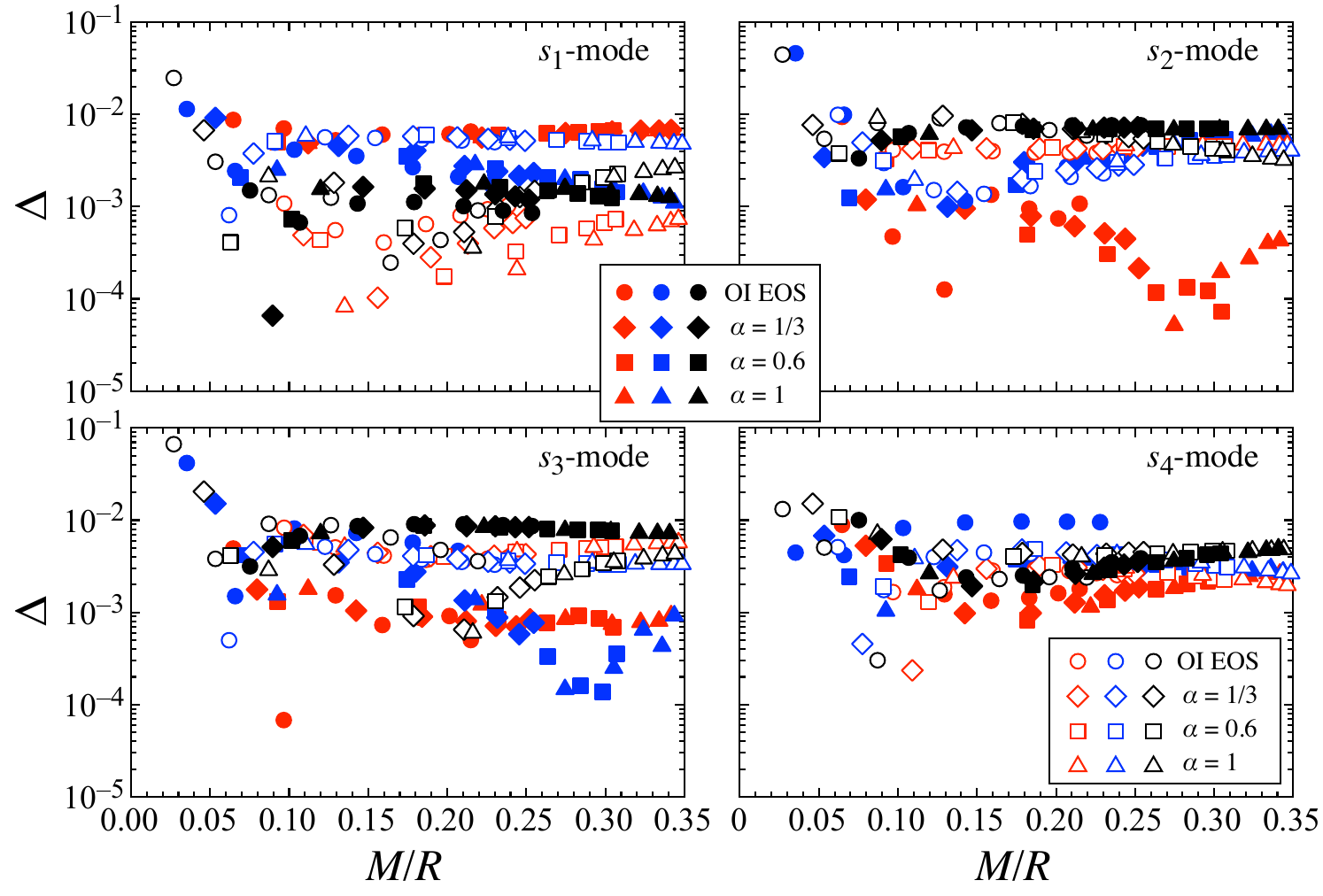} 
\end{center}
\caption{
For the $s_i$-mode frequencies with $i=1-4$, the relative deviation of the frequencies determined by solving the eigenvalue problem from those evaluated with the fitting formulae given by Eqs.~(\ref{eq:sR}), (\ref{eq:a0i}), and (\ref{eq:a1i}) are shown as a function of the stellar compactness, $M/R$, for the various stellar models constructed with different saturation parameters and stiffness of the core EOS. The circles, diamonds, squares, and triangles denote the different stiffness of the core EOS, i.e., original OI-EOS (circles), $\alpha=1/3$ (diamonds), $\alpha=0.6$ (squares), and $\alpha=1$ (triangles). The different colors and whether marks are open or filled denote the EOS models with different saturation parameters.
Taken from Ref.~\cite{Sotani24b}. }
\label{fig:delta_s}
\end{figure}

\begin{figure}[tbp]
\begin{center}
\includegraphics[scale=0.5]{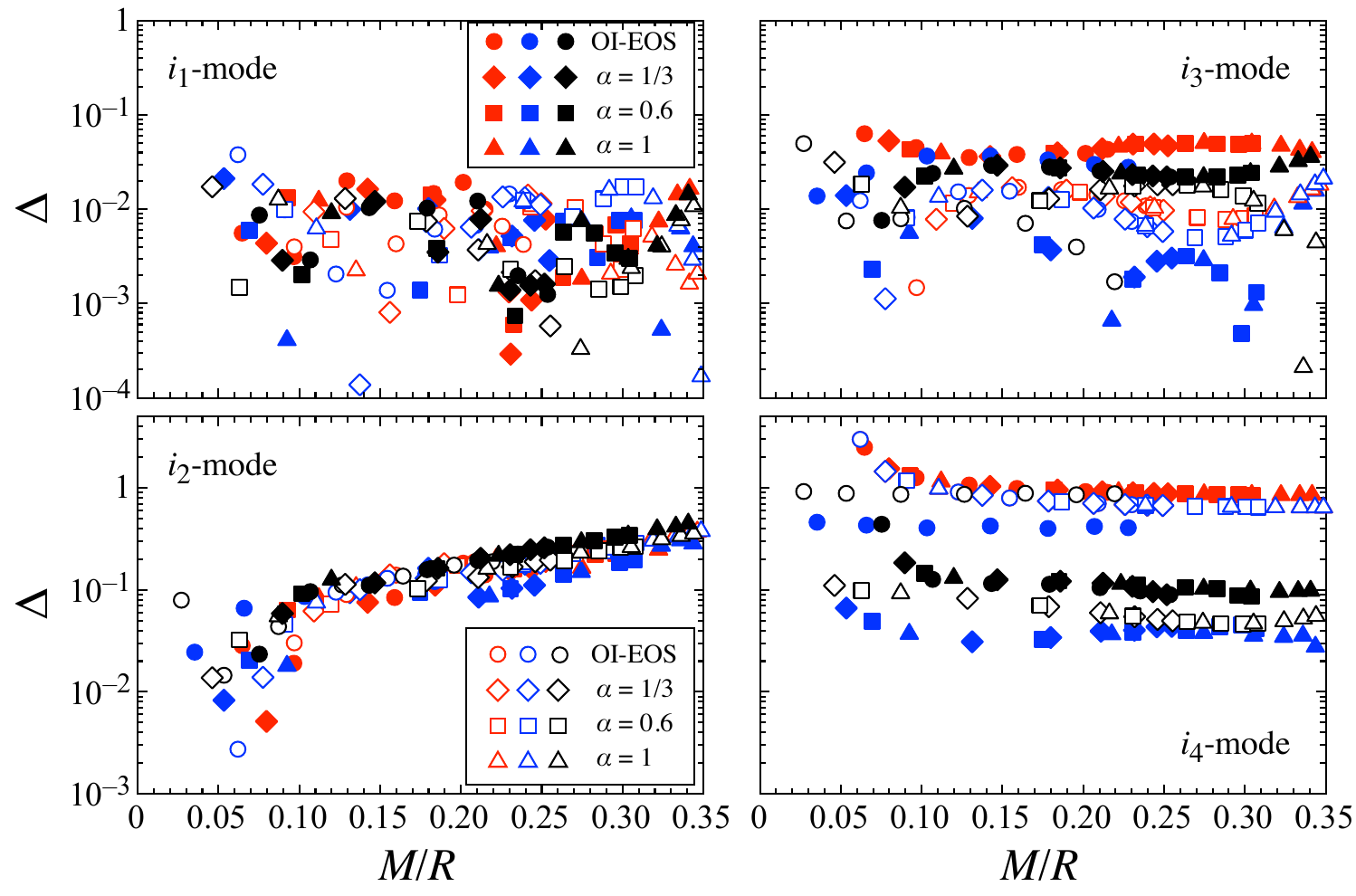} 
\end{center}
\caption{
For the $i_i$-mode frequencies with $i=1-4$, the relative deviation, $\Delta$, using the fitting formulae given by Eqs.~(\ref{eq:iM}), (\ref{eq:bji}), and (\ref{eq:bj4}), is shown as a function of $M/R$. The meaning of the marks is the same as in Fig.~\ref{fig:delta_s}. Taken from Ref.~\cite{Sotani24b}. }
\label{fig:delta_i}
\end{figure}

To check how well our fitting formulae work, we calculate the relative deviation, $\Delta$, of the frequencies from those evaluated with the fitting formulae, as
\begin{equation}
  \Delta \equiv \frac{|f-f_{\rm em}|}{f}, \label{eq:Delta}
\end{equation}
where $f$ and $f_{\rm em}$ respectively denote the frequencies determined by numerically solving the eigenvalue problem and those evaluated with the fitting formulae given by Eqs.~(\ref{eq:sR}), (\ref{eq:a0i}), and (\ref{eq:a1i}) for the $s_i$-modes and Eqs.~(\ref{eq:iM}), (\ref{eq:bji}), and (\ref{eq:bj4}) for the $i_i$-modes, and show them in Fig.~\ref{fig:delta_s} for the $s_i$-modes and in Fig.~\ref{fig:delta_i} for the $i_i$-modes, respectively. From Fig.~\ref{fig:delta_s}, we find that our fitting formulae work well for the $s_i$-mode frequencies with less than $\sim 1\%$ error, except for the unrealistic low-mass neutron stars. Meanwhile, as shown in Fig.~\ref{fig:delta_i}, using our fitting formulae, the $i_1$- and $i_3$-mode frequencies can be evaluated within a few $\%$ and $\sim 10\%$ accuracy, respectively, but the $i_2$- and $i_4$-mode frequencies are evaluated only an order of magnitude due to the weak correlations with nuclear saturation parameters. We note that we derive the fitting formulae for the $s$- and $i$-mode frequencies using the specific combination of $K_0$ and $L$ listed in Table~\ref{tab:yzi}, but this is not unique, and better-fitting formulae may exist.

\subsection{Accuracy of the Cowling approximation}
\label{sec:Cowling}

The above discussion has been done, using the $s$- and $i$-mode frequencies determined with the Cowling approximation. In this subsection, we discuss the accuracy of this approximation in the $s$- and $i$-mode frequencies by comparing the frequencies determined via the eigenvalue problem with the metric perturbations to those with the Cowling approximation~\cite{KS25}. As a result, we find that the $s$-mode frequencies with the Cowling approximation surprisingly agree well with those with the metric perturbations within $1\%$ accuracy, which may come from numerical accuracy and/or marginal differences in the implementations of the numerical codes in the two approaches. In addition, we find that the eigenfunctions of the $s$-modes determined with the metric perturbations almost overlap with those with the Cowling approximation (see Figs.~2 and 3 in Ref.~\cite{KS25}). That is, the frequencies and the corresponding eigenfunctions of $s$-modes virtually receive no impact from the Cowling approximation. Considering that the Cowling approximation overestimates the $f$-mode frequencies $\sim 20\%$~\cite{MK97}, this is an astonishing result. In contrast, we also find that the $i$-mode frequencies determined with the Cowling approximation deviate from those with the metric perturbations by about $20\%$, e.g., the Cowling approximation overestimates the $i_1$-mode frequency and underestimates the $i_2$-mode frequency, compared to the frequencies determined with the metric perturbations~\cite{KS25}. One also observes deviations in the corresponding eigenfunctions. Such a discrepancy in the frequencies as well as the corresponding eigenfunctions of the $i$-modes is somewhat unexpected. As the $i$-modes are very weakly coupled to the spacetime and hardly radiate gravitational waves (see the next paragraph), we would expect them to be almost unaffected by the Cowling approximation, as for the $s$-modes. While this result may be physical in nature, if it is not, we were ultimately unable to identify the cause of this discrepancy.
Anyway, the deviation of the $i$-mode frequencies determined with the metric perturbations from those with the Cowling approximation is much larger than the uncertainties of the fitting formulae for the $i_1$-mode frequencies derived from the calculation with the Cowling approximation. In other words, to extract the physical properties under realistic situations, it is necessary to calculate the frequency of the $i$ mode using metric perturbations and to rederive the fitting formula, even if our fitting formula for the $i_1$-mode frequencies derived with the Cowling approximation qualitatively represents specific characteristics of the $i_1$-mode (although this point itself requires verification).

As an advantage of a full relativistic approach taking into account the metric perturbations, one can discuss the damping rate of the gravitational waves. Because gravitational waves carry away energy, the stellar oscillations would damp out; i.e., the eigenfrequencies become complex, the so-called quasi-normal modes. Through a standard numerical method, we tried to determine the damping rate for the $s$- and $i$-mode oscillations, but the damping rate is too low to calculate, which is comparable to or less than the numerical accuracy. Therefore, to estimate the damping rate, we adopt two different approaches, i.e., one is the method using the ratio of the outgoing and ingoing wave amplitudes for purely real-valued frequencies~\cite{Nils02}, based on the standard Breit-Wigner resonance, and the other is the well-known quadrupole formula~\cite{Kip69b,LS18}, using the estimation of the oscillation energy with the specific mode~\cite{LS18,Kip69a}. Then, we could estimate the damping time to be $\gtrsim 10^6$ sec. for the $s_1$-modes and $> 10^{8}$ sec. for the $i_1$-modes. Thus, the damping rate of the $s$- and $i$-modes is very low, which means that the gravitational wave emissions from these modes are so weak as to be virtually negligible.
We emphasize that the damping time estimated here is due to gravitational radiation alone, not necessarily the physical lifetimes of the modes, which may be affected by other dissipative or nonlinear processes.

\section{Conclusion}
\label{sec:conclusion}

In this article, we synthesized the results of Refs.~\cite{Sotani23,Sotani24b,KS25} and discussed their implications for NS seismology.
The $s$- and $i$-mode frequencies excited by the elasticity inside the neutron star are directly associated with the properties in the elastic region. In this study, we focus on the crust elasticity, even though there may exist another elastic region inside the star. The $s$- and $i$-mode frequencies might be observed in the precursor signal resulting from the resonant shattering that occurred before the main short gamma-ray burst corresponding to the merger of a binary neutron star. If observed, this would be a good opportunity to detect the gravitational waves from the neutron stars, where one could extract the neutron star properties, especially the crust properties, through the technique of asteroseismology. Nevertheless, as the excited frequencies strongly depend on uncertainties, such as neutron star mass, radius, and EOS, it may be difficult to directly obtain the physical information from the observed frequencies of gravitational waves. To solve this difficulty, we could derive the fitting formulae for the $s_i$-mode with $i=1-4$ and the $i_1$-mode frequencies as a function of stellar compactness and the combination of the nuclear saturation parameters, $K_0$ and $L$, independently of the stiffness of the core EOS, at least when the EOS in a higher-density region can be characterized only by the specific constant sound velocity. 
We have to emphasize that we adopt the barotropic approximation in this manuscript; i.e., a realistic EOS with stratification may change the mode identification itself.

Our fitting formula for the elastic frequencies becomes a function of neutron star properties, i.e., the mass $M$ and radius $R$, and the EOS parameters, $K_0$ and $L$, indicating that the observation of a single frequency does not directly provide a constraint on $K_0$ and $L$ without additional information about the stellar parameters $M$ and $R$. Even if $M$ and $R$ were independently known, one frequency would generally constrain a combination of the nuclear parameters rather than determine them separately. The possibility of obtaining stronger constraints from several observed modes depends on the independence of the corresponding parameter combinations, their measurement uncertainties, and the reliability of mode identification. This means that the fitting formula derived in this article provides one of the relationships in the parameter space comprising the EOS parameters and the properties of the star; by combining this with relationships obtained from entirely different observations, it is possible to narrow down the parameter space. In such a way, one of the potential sources of the elastic oscillations considered in this article is crustal oscillations, which may emerge as a precursor before a binary neutron star merger observed as a short gamma-ray burst.   
Resolving these oscillation modes considered in this article is currently very difficult in observations as well as in numerical simulations. Nevertheless, should the resolution of both numerical simulations and observations improve dramatically in the future, and should frequencies arising from crustal elasticity---such as those discussed here---be observed, it may be possible to shed light on properties of the crust that remain unclear at present.

Since we consider EOS models (OI-EOS), which are characterized only by the nuclear saturation parameters, i.e., the lowest-order expansion coefficients of zero-temperature homogeneous nuclear matter in the vicinity of the saturation density of symmetric nuclear matter, our statement in this article is valid only when the correlations with the nuclear saturation parameters are well preserved. If not, additional analysis is needed to validate our statement. A useful way of verifying the robustness of our findings is to examine whether various kernel functions such as more flexible nucleonic EOS models proposed in Ref.~\cite{Montefusco26}---with roughly the same saturation parameter but differing in their remaining parameters---predict equivalent elastic mode frequencies. Such an extension will be done somewhere in the future. 
In addition, we also have to examine the effects neglected in this article, such as neutron superfluidity and composition fraction inside the star, and/or the case without simple approximations, such as the zero shear modulus inside the slab-like nuclei.
Actually, the frequencies of elastic oscillations possibly increase due to neutron superfluidity, which depends on the effective dynamical coupling between the dripped neutrons and the crust, including entrainment.



\section{Patents}

\funding{This work is supported in part by the Mitsubishi Foundation through grant No. 202510029.
}

\institutionalreview{Not applicable.
}

\informedconsent{Not applicable.
}

\dataavailability{Dataset available on request from the authors.
} 



\acknowledgments{We are grateful to Christian J. Kr\"{u}ger  for valuable comments. 
}

\conflictsofinterest{The authors declare no conflicts of interest. The funders had no role in the design of the study; in the collection, analyses, or interpretation of data; in the writing of the manuscript; or in the decision to publish the results.
} 



\abbreviations{Abbreviations}{
The following abbreviations are used in this manuscript:\\

\noindent 
\begin{tabular}{@{}ll}
EOS & Equation of state\\
TOV & Tolman-Oppenheimer-Volkoff
\end{tabular}
}

\appendixtitles{yes} 
\appendixstart
\appendix


\section[\appendixname~\thesection]{Mass-radius relations}
\label{sec:appendix_A}

\begin{figure}[tbp]
\begin{center}
\includegraphics[scale=0.5]{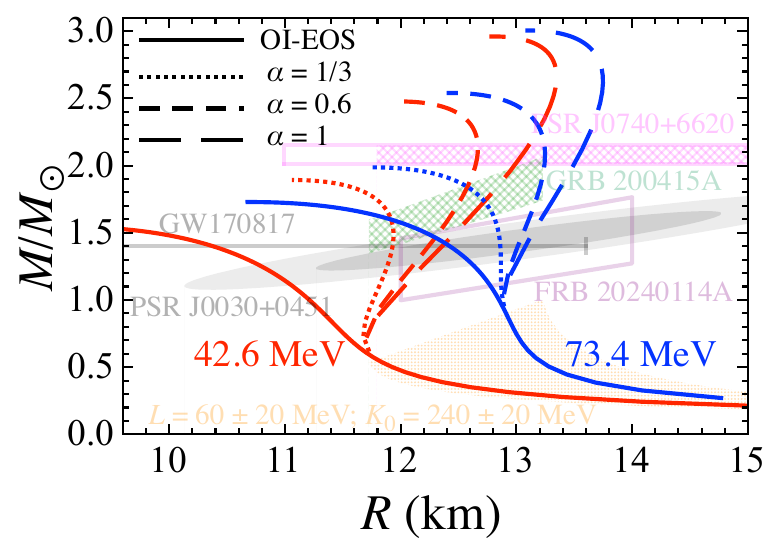} 
\end{center}
\caption{
Mass-radius relations of the neutron star models considered in Fig.~\ref{fig:ismode}. The solid lines denote the stellar models constructed with the OI-EOS, while the dashed and dotted lines denote those constructed with the OI-EOS connected to the EOS given by Eq. (\ref{eq:eos_high}) in a higher density region. For reference, some of the observational constraints are also shown (see the text for the details). 
 }
\label{fig:MR}
\end{figure}

In this appendix, we show the mass-radius relations of the neutron star models considered in Fig.~\ref{fig:ismode}. In Fig.~\ref{fig:MR}, the mass-radius relations for the neutron star models are shown, where the solid lines denote the stellar models constructed with the OI-EOS, while the dashed and dotted lines denote those constructed with the OI-EOS connected to the EOS given by Eq. (\ref{eq:eos_high}) in a higher density region. For reference, astronomical constraints are also shown, i.e., NICER observations for PSR J0030+0451~\cite{Riley19,Miller19} and for PSR J0740+6620~\cite{Riley21,Miller21,Dittmann24}; the gravitational waves from binary neutron star merger, GW170817~\cite{Annala18}; the high-frequency QPOs from GRB 200415A~\cite{SKS23}; and quasi-periodic oscillaitons from FRB 2024114A~\cite{Sotani26b}. In addition to the astronomical constraints, the neutron star models, whose central density is low, can be constrained from the experimental values of $K_0$ and $L$ using the low-mass neutron star mass formula proposed in Ref.~\cite{SIKO14}. The right-bottom shaded region is such an allowed region by adopting the fiducial values of $K_0=240\pm 20$ MeV~\cite{Shlomo06} and $L=60\pm 20$ MeV~\cite{Vinas14,Li19}. From this figure, one can observe that the stellar models considered in Fig.~\ref{fig:ismode} cover a surprisingly wide range of the mass-radius plane.

\section[\appendixname~\thesection]{Dependence of the transition density between an elastic crust and the envelope.}
\label{sec:appendix_B}

In this study, we assume that the transition density between the surface of the spherical nuclei and the envelope is $10^{10}$ g/cm$^3$. However, if this transition density is different from this assumed value, the frequencies of the elastic oscillations, especially the $i$-mode frequencies, may change, depending on the transition density. In this appendix, we examine how the $i$-mode frequencies shift as the transition density varies. For this purpose, we simply consider the stellar models with an elastic region inside the spherical and cylindrical nuclei, while the elasticity inside the slab-like, cylindrical-hole, and spherical-hole is neglected. That is, the number of excited $i$-modes is only two, which can be excited at the interface between the surface of the spherical nuclei phase and the envelope, and between the cylindrical and slab-like nuclei phases. Here, the $i_1$-mode is the oscillation mode whose amplitude becomes large at both boundaries, while the $i_2$-mode is the mode whose amplitude mainly becomes large at the boundary between the crust and the envelope. 

\begin{figure}[tbp]
\begin{center}
\includegraphics[scale=0.5]{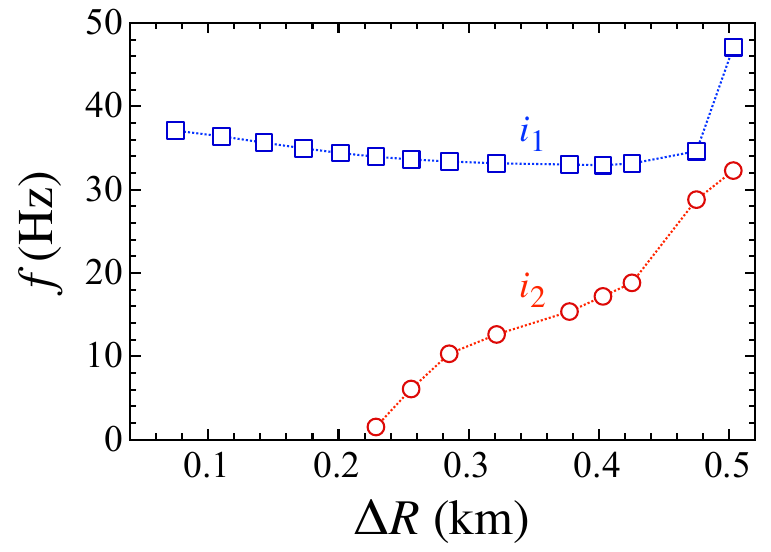} 
\end{center}
\caption{
The dependence of the $i_1$- and $i_2$-mode frequencies on the crust thickness (or the transition density between the surface of the spherical nuclei phase and the envelope. In particular, for simplicity, the neutron star models are considered with an elastic region inside the spherical and cylindrical nuclei, where the elasticity in the slab-like, cylindrical-hole, and spherical-hole nuclei is neglected. That is the totally the two $i$-modes can be excited at the interface between the surface of phase of spherical nuclei and envelope, and between the phase of cylindrical and slab-like nuclei. The neutron star model shown here is $1.4M_\odot$ and 12.4 km, using the OI-EOS with $L=73.4$ MeV and $K_0=230$ MeV. Taken from Ref.~\cite{Sotani23}. }
\label{fig:Dep_dR}
\end{figure}

Then, the transition density between the surface of the spherical nuclei and the envelope varies from $10^{12}$ to $10^{14}$ g/cm$^3$, focusing on the stellar model with $1.4M_\odot$ and 12.4 km constructed with the OI-EOS with $L=73.4$ MeV and $K_0=230$ MeV. The corresponding change in the crust thickness is from
$\Delta R=0.503$ to $0.075$ km. The behavior of the $i_1$- and $i_2$-mode frequencies is shown in Fig.~\ref{fig:Dep_dR}. From this figure, one can observe that $i_1$-mode frequencies are almost independent of the variation of the transition density between the crust and the envelope, while the $i_2$-mode frequencies significantly depend on it. In fact, the $i_2$-mode frequencies decrease and eventually disappear when the transition density increases, corresponding to the decrease of the crust thickness. In a realistic (cold) neutron star, the transition density may not be so high that we can observe a significant shift in the $i$-mode frequencies, but it may happen in a neutron star where the crust is just forming during neutron star cooling.

\begin{adjustwidth}{-\extralength}{0cm}

\reftitle{References}

\PublishersNote{}
\end{adjustwidth}
\end{document}